# Explainable artificial intelligence (XAI) in deep learning-based medical image analysis

Bas H.M. van der Velden, Hugo J. Kuijf, Kenneth G.A. Gilhuijs, Max A. Viergever

*Image Sciences Institute, University Medical Center Utrecht, Utrecht, The Netherlands*

**Corresponding author:**

Bas van der Velden

Fax: NA

Phone: +31 88 75 57772

Email address: bvelden2@umcutrecht.nl

Postal address: Image Sciences Institute, University Medical Center Utrecht, Utrecht University, Q.02.4.45, P.O. Box 85500, 3508 GA Utrecht, The Netherlands

**Abstract**

With an increase in deep learning-based methods, the call for explainability of such methods grows, especially in high-stakes decision making areas such as medical image analysis. This survey presents an overview of eXplainable Artificial Intelligence (XAI) used in deep learning-based medical image analysis. A framework of XAI criteria is introduced to classify deep learning-based medical image analysis methods. Papers on XAI techniques in medical image analysis are then surveyed and categorized according to the framework and according to anatomical location. The paper concludes with an outlook of future opportunities for XAI in medical image analysis.

# 1. Introduction

Deep learning has invoked tremendous progress in automated image analysis. Before that, image analysis was commonly performed using systems fully designed by human domain experts. For example, such image analysis system could consist of a statistical classifier that used handcrafted properties of an image (i.e., features) to perform a certain task. Features included low-level image properties such as edges or corners, but also higher-level image properties such as the spiculated border of a cancer. In deep learning, these features are learned by a neural network (in contrast to being handcrafted) to optimally give a result (or output) given an input. An example of a deep learning system could be the output 'cancer' given the input of an image showing a cancer.

Neural networks typically consist of many layers connected via many nonlinear intertwined relations. Even if one is to inspect all these layers and describe their relations, it is unfeasibly to fully comprehend how the neural network came to its decision. Therefore, deep learning is often considered a 'black box'. Concern is mounting in various fields of application that these black boxes may be biased in some way, and that such bias goes unnoticed. Especially in medical applications, this can have far-reaching consequences.

There has been a call for approaches to better understand the black box. Such approaches are commonly referred to as interpretable deep learning or eXplainable Artificial Intelligence (XAI) (Adadi and Berrada (2018); Murdoch et al. (2019)). These terms are commonly interchanged; we will use the term XAI. Some notable XAI initiatives include those from the United States Defense Advanced Research Projects Agency (DARPA), and the conferences on Fairness, Accountability, and Transparency by the Association for Computing Machinery (ACM FAccT).

The stakes of medical decision making are often high. Not surprisingly, medical experts have voiced their concern about the black box nature of deep learning (Jia et al. (2020)), which is the current state of the art in medical image analysis (Litjens et al. (2017); Meijering (2020); Shen et al. (2017)). Furthermore, regulations such as the European Union's General Data Protection Regulation (GDPR, Article 15) require the right of patients to receive meaningful information about how a decision was rendered.

Researchers in medical imaging are increasingly using XAI to obtain insight into their algorithms. In this survey, we aim to give a comprehensive overview of papers using XAI in medical image analysis. We chose to focus solely on papers that used deep learning-based XAI in medical image analysis. The search strategy for inclusion of papers is detailed in Appendix 1. In short, it followed a systemic review procedure, discussion with colleagues, and a snowballing approach – investigating papers referenced by the included papers and papers that refer to the included papers, to come to the final list of surveyed articles.

The survey is structured as follows: We will first introduce the taxonomy of XAI and describe a framework to classify XAI techniques in Section 2. In Section 3, the discussed papers are characterized according to this XAI framework. We will discuss applications of XAI techniques in medical image analysis. In case of multiple papers using the same technique, we will discuss some early adopters and summarize the rest of the papers in the tables. Since XAI techniques often originate from computer vision, we will elaborate on papers that adapted XAI techniques from computer vision by adding domain knowledge from the medical imaging field. The papers are grouped in the tables according to explanation method and according to anatomical location. This survey adds to the review of Reyes et al. (2020); since they mainly discussed techniques in computer vision, without extensively evaluating the adaptation of such techniques throughout medical image analysis. Furthermore, we describe if and how techniques from computer vision have been adapted specifically for medical image analysis. This survey adds to the review of Huff et al. (2021), since they mostly focused on examples of visual explanation, while our survey aims for a more holistic approach including non-visual explanation, critiques on XAI, and methods for evaluating XAI. Additionally, we systematically survey papers, reflecting the current status of the field of XAI in medical imaging. The survey is concluded in Section 4 by discussing the state of the art of XAI in medical image analysis and an outlook of the opportunities of XAI.

## 2. Explainable Artificial Intelligence (XAI) framework

In this section, we will give a brief overview of Explainable Artificial Intelligence (XAI) techniques found in deep learning for medical image analysis. For exhaustive surveys focused solely on XAI, please refer to Adadi and Berrada (2018) and Murdoch et al. (2019).

We will distinguish XAI techniques based on three criteria: model-based versus post hoc, model-specific versus model-agnostic, and global versus local (i.e., the scope of the explanation). The framework of these three criteria is adapted from the surveys of Adadi and Berrada (2018) and Murdoch et al. (2019) and is depicted in Figure 1. The following paragraphs will describe these criteria.

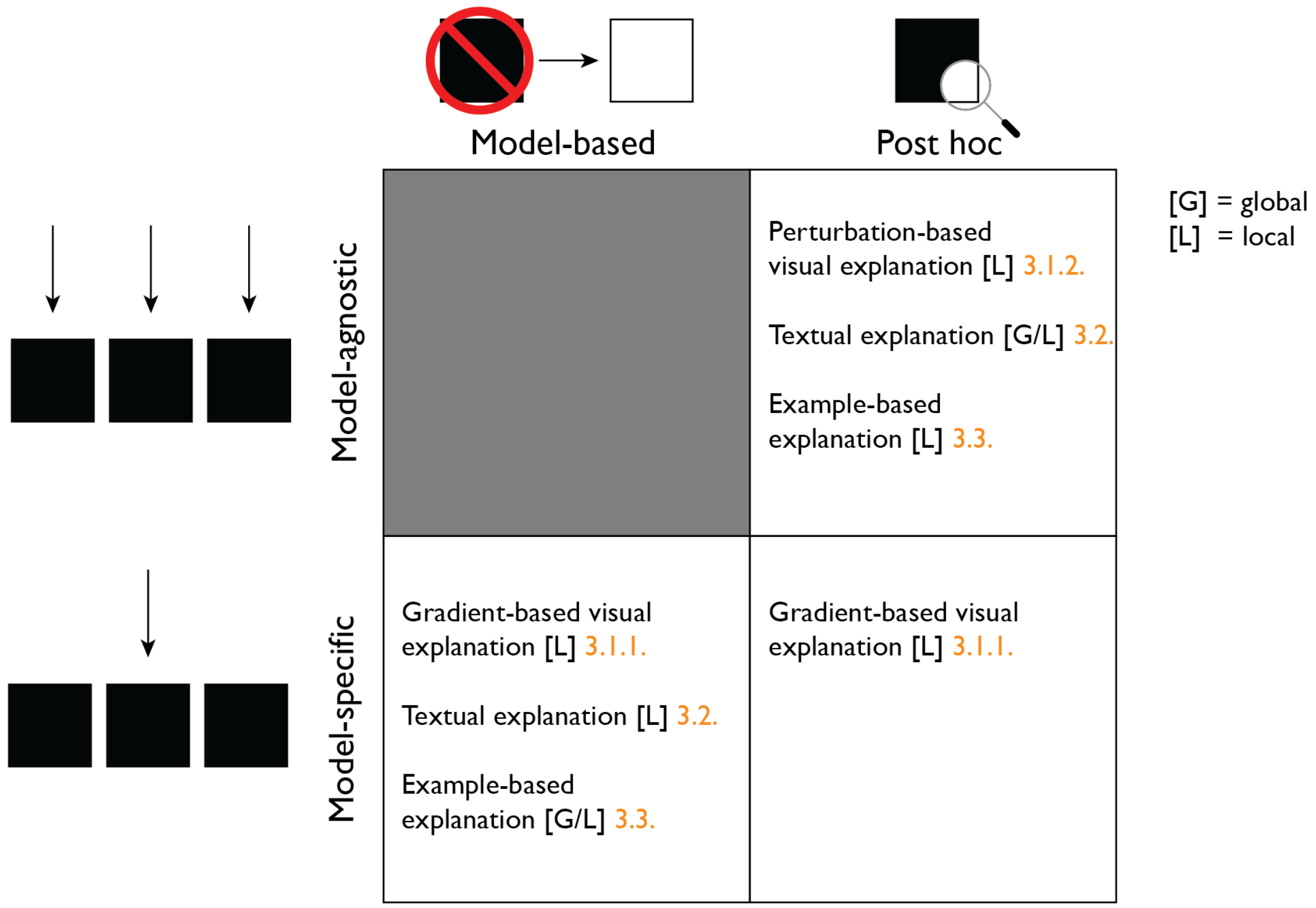

**Figure 1:** The eXplainable Artificial Intelligence (XAI) framework proposed in this paper. A rough overview of XAI techniques (discussed in Section 3) is classified according to this framework. The orange number refers to the section number in the manuscript where the XAI technique is described.

### 2.1. Model-based versus post hoc explanation

The first distinction we make is model-based explanation versus post hoc explanation (Figure 1). In deep learning, the model generally refers to the neural network, and we will use the terms model and neural network interchangeable throughout this survey.

#### 2.1.1. Model-based explanation

Model-based explanation refers to models, e.g. a linear regression model or a support vector machine, that are simple enough to be understood, but sophisticated enough to fit a relationship between input and output well (Murdoch et al. (2019)). These are often the traditional machine learning models. Examples of model-based explanation enforce the use of a limited amount of features (i.e., sparsity), or enforce a human to be able to internally reason about the model's entire decision-making process (i.e., simulatability) (Murdoch et al. (2019)). For example, models that enforce sparsity such as the least absolute shrinkage and selection operator (LASSO, Tibshirani (1996)), force many coefficients to zero. Hence, a select subset of features leads to an output, making the inner construct of this model explainable.

Since the focus of our survey is on XAI methods for deep learning, model-based explanation by enforcing sparsity or simulatability is infeasible. The model in deep learning is a deep neural network, typically with thousands to millions of weights, which is neither sparse, nor suited for a human to internally simulate and reason about the models entire decision making. However, one of the methods mentioned by Murdoch et al. (2019) was model-based feature engineering, i.e., automated approaches for constructing explainable features.

### 2.1.2. Post hoc explanation

Analyzing a trained model (i.e., a neural network in deep learning) to achieve insight into learned relationships is referred to as post hoc explanation. An important distinction between post hoc explanation and model-based explanation is that the former trains a neural network and subsequently attempts to explain the behavior of the ensuing black box network, whereas the latter forces the model to be explainable.

Methods that provide post hoc explanation include inspection of learned features, feature importance, and interaction of features (Abbasi-Asl and Yu (2017); Olden et al. (2004); Tsang et al. (2018)); as well as visual explanation by saliency maps (Selvaraju et al. (2017); Simonyan et al. (2013); Springenberg et al. (2014); Zeiler and Fergus (2014); Zhou et al. (2016)).

## 2.2. Model-specific versus model-agnostic explanation

The distinction between model-specific and model-agnostic explanation is related to that between model-based and post hoc explanation (Adadi and Berrada (2018)), but there are some nuanced differences.

### 2.2.1. Model-specific explanation

Model-specific explanation methods are limited to particular classes of models. For example, such a method may use attributes that are specific to a type of neural network. A drawback is that by aiming at model-specific explanation, we limit our choice of model, thereby potentially excluding a model that could better fit the output to the input data.

Model-based explanation is by definition model-specific (Adadi and Berrada (2018)), but model-specific explanation is not necessary model-based. Some post hoc saliency mapping techniques are examples of techniques that are specific to a certain class of convolutional neural networks (CNNs), but are not model-based explanation methods (Murdoch et al. (2019)).

#### 2.2.2. Model-agnostic explanation

Model-agnostic explanation is independent of the model choice, operating solely on the input and the output of the model. By perturbing the input of a model, the user can inspect what the change is in the output of the model. This can therefore explain which regions are driving the output of the model. Model-agnostic explanation is naturally post hoc.

### 2.3. Scope of explanation

The scope of an explanation distinguishes between explanation for an entire model (global) versus explanation for a single output (local).

#### 2.3.1. Global explanation

Global explanation, also called dataset-level explanation, provides general relationships learned by the model. For example, global explanation could provide feature importance scores at the dataset level, i.e., how much do features contribute to the output across the entire dataset (Olden et al. (2004)). As an illustration, one might observe from a model that – or even how much – high blood pressure increases the risk of a cardiac event. Another example of global explanation could be visualization of learned filters, i.e., which features are extracted by the neural network and to what extent are they meaningful to the task at hand (Olah et al. (2017); Zeiler and Fergus (2014)).

### 2.3.2. Local explanation

Local explanation provides explanation of a single input. In the example of cardiac risk, an input would be a single person. Local explanation would therefore explain why blood pressure is important to the risk of cardiac event for that single person, whereas global explanation would describe the relation of blood pressure with risk of cardiac events across the entire dataset. Another example of a local explanation could be a saliency map pinpointing to a brain tumor on magnetic resonance imaging (MRI) to explain which part of the MRI mainly contributed to the classifier output ‘tumor’. Since this explains which part of the image drives the classifier to its output ‘tumor’ for that single person, this is a local explanation.

## 3. XAI in medical image analysis

In this section, we will present which XAI techniques are used in medical image analysis, and we will discuss adaptations of the methods typically seen in computer vision. We categorize the explanation methods into three types: visual, textual, and example-based; and we will classify each method according to the framework of model-based versus post hoc, model-specific versus model-agnostic, and global versus local explanation (Figure 1). Table 1 provides an overview of the most frequently used techniques, classified according to the definitions of Section 2.

**Table 1:** Overview of eXplainable AI (XAI) techniques used in medical image analysis, classified by the framework from Section 2.

| Technique | Section | Authors | Model-based | Post hoc | Model-specific | Model-agnostic | Global | Local |
|---|---|---|---|---|---|---|---|---|
| **Visual explanation** | **3.1.** | | | | | | | |
| *Backpropagation-based approaches* | 3.1.1 | | | | | | | |
| Backpropagation | 3.1.1.1. | Simonyan et al. (2013) | | ✓ | ✓ | | | ✓ |
| Deconvolution | 3.1.1.1. | Zeiler and Fergus (2014) | | ✓ | ✓ | | | ✓ |
| Guided backpropagation | 3.1.1.1. | Springenberg et al. (2014) | | ✓ | ✓ | | | ✓ |
| Class activation mapping (CAM) | 3.1.1.2. | Zhou et al. (2016) | | ✓ | ✓ | | | ✓ |
| Gradient-weighted class activation mapping (Grad-CAM) | 3.1.1.3. | Selvaraju et al. (2017) | | ✓ | ✓ | | | ✓ |
| Layer-wise relevance propagation (LRP) | 3.1.1.4. | Bach et al. (2015) | | ✓ | ✓ | | | ✓ |
| Deep SHapley Additive exPlanations (Deep SHAP) | 3.1.1.5. | Lundberg and Lee (2017) | | ✓ | ✓ | ✓* | ✓* | ✓ |
| Trainable attention | 3.1.1.6. | Jetley et al. (2018) | ✓ | | ✓ | | | ✓ |
| *Perturbation-based approaches* | 3.1.2 | | | | | | | |
| Occlusion sensitivity | 3.1.2.1. | Zeiler and Fergus (2014) | | ✓ | | ✓ | | ✓ |
| Local Interpretable Model-agnostic Explanations (LIME) | 3.1.2.2. | Ribeiro et al. (2016) | | ✓ | | ✓ | | ✓ |
| Meaningful Perturbation | 3.1.2.3. | Fong and Vedaldi (2017) | | ✓ | | ✓ | | ✓ |
| Prediction difference analysis | 3.1.2.4. | Zintgraf et al. (2017) | | ✓ | | ✓ | | ✓ |
| **Textual explanation** | **3.2.** | | | | | | | |
| Image captioning | 3.2.1. | Vinyals et al. (2015) | ✓ | | ✓ | | | ✓ |
| Image captioning with visual explanation | 3.2.2. | Zhang et al. (2017a) | ✓ | | ✓ | | | ✓ |
| Testing with Concept Activation Vectors (TCAV) | 3.2.3. | Kim et al. (2018) | | ✓ | | ✓ | ✓ | ✓ |
| **Example-based explanation** | **3.3.** | | | | | | | |
| Triplet networks | 3.3.1. | Hoffer and Ailon (2015) | ✓ | | ✓ | | ✓ | ✓ |
| Influence functions | 3.3.2. | Wei Koh and Liang (2017) | | ✓ | | ✓ | ✓ | |
| Prototypes | 3.3.3 | C. Chen et al. (2019) | ✓ | | ✓ | | | ✓ |

* Deep Shapley Additive exPlanations are post hoc and model-specific because of the optimization method, but Shapley Additive exPlanations can also be global and model-agnostic.

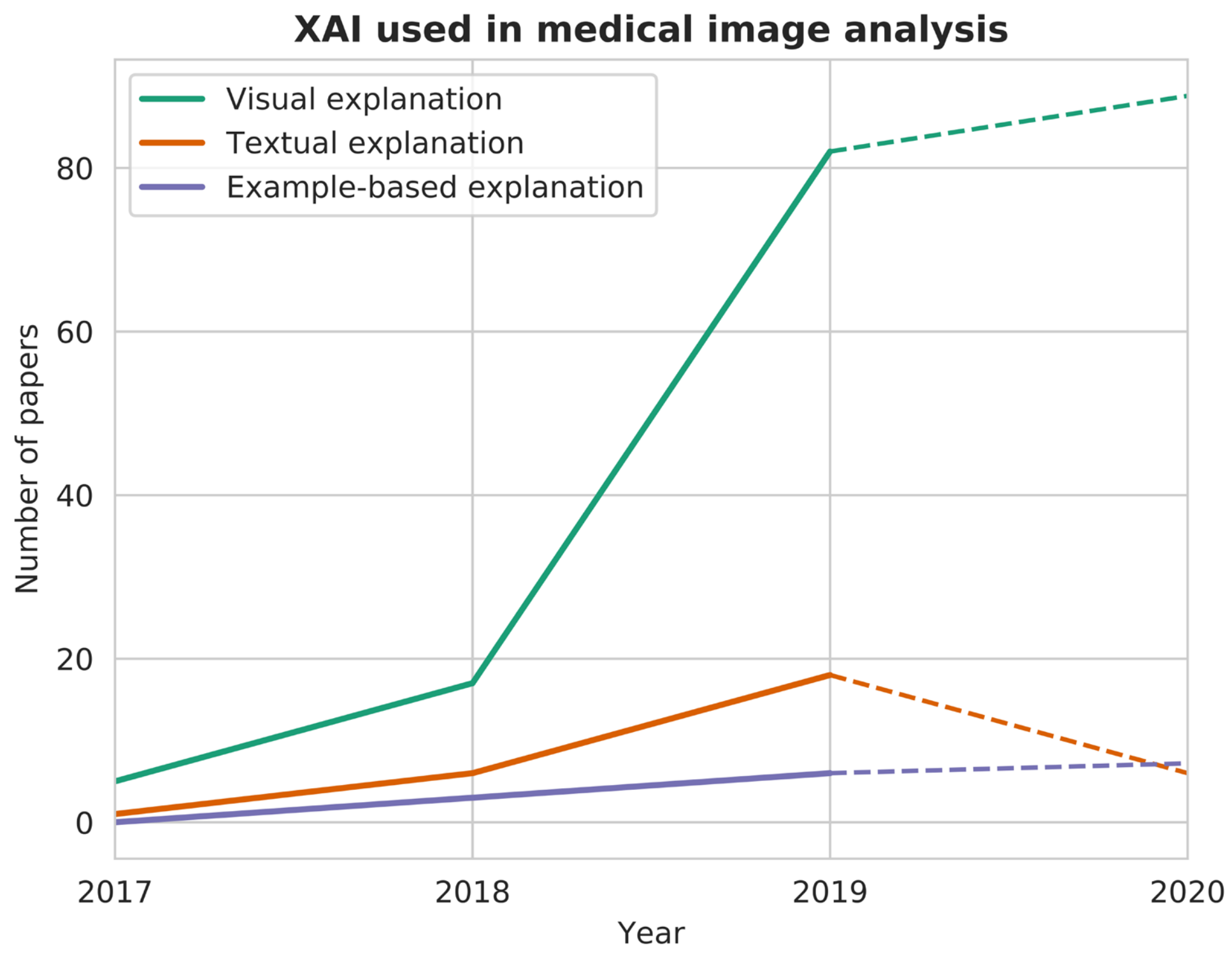

**Figure 2:** Number of papers published per year in medical image analysis, for the three types of XAI techniques. Most papers use a visual explanation. The y-axis shows the number of papers included in this survey, the x-axis shows the year these papers were published in. The dashed line for 2020 is an extrapolation given the situation on October 31, 2020.

### 3.1. Visual explanation

Visual explanation, also called saliency mapping, is the most common form of XAI in medical image analysis. Saliency maps show the important parts of an image for a decision. Most saliency mapping techniques use backpropagation-based approaches, but some use perturbation-based or multiple instance learning-based approaches. These approaches will be discussed below. An overview of papers using saliency maps in medical imaging is shown in Table 2.

### 3.1.1. Backpropagation-based approaches

#### *3.1.1.1. (Guided) backpropagation and deconvolution*

Some of the earliest techniques to create saliency maps highlighted pixels that had the highest impact on the analysis output. Examples included visualization of partial derivatives of the output on pixel level (Simonyan et al. (2013)), deconvolution (Zeiler and Fergus (2014)), and guided backpropagation (Springenberg et al. (2014)). These techniques provided local, model-specific (only for CNNs), post hoc explanation. These techniques have been used in medical image analysis. For example, de Vos et al. (2019) estimated the amount of coronary artery calcium per cardiac or chest computed tomography (CT) image slice, and used deconvolution to visualize from where in the slice the decision was based on.

#### *3.1.1.2. Class Activation Mapping (CAM)*

Zhou et al. (2016) introduced Class Activation Mapping (CAM). They replaced the fully connected layers at the end of a CNN by global average pooling on the last convolutional feature maps. The class activation map was a weighted linear sum of presence of visual patterns (captured by the filters) at different spatial locations. This technique provided local, model-specific, post hoc explanation. Several researchers used this technique in medical imaging (Table 2).

CAMs have also been used in medical image analysis in ensembles of CNNs. For example, Jiang et al. (2019) constructed an ensemble of Inception-V3, ResNet-152, and Inception-ResNet-V2 to distinguish fundus images of healthy subjects or patients with mild diabetic retinopathy from those with moderate or severe diabetic retinopathy; and provided a weighted combination of the resulting CAMs for localization of diabetic retinopathy. Lee et al. (2019b) constructed CAMs of the output of an ensemble of four CNNs:

VGG-16, ResNet-50, Inception-V3, and Inception-ResNet-V2, for the detection of acute intracranial hemorrhage.

Since medical images often contain information at multiple scales, multi-scale CAMs have also been proposed. Liao et al. (2019) concatenated feature maps at three scales which were subsequently provided as input for the global average pooling. The provided activation maps showed higher resolution than single-scale maps, and were better at identifying small structures on fundus images of the retina. Shinde et al. (2019a) concatenated the feature maps of each layer before max-pooling and also gave those as input to a global average pooling layer. Their 'High Resolution' CAMs provided accurate localizations of brain tumors on MRI. García-Peraza-Herrera et al. (2020) proposed extracting CAMs at multiple resolutions. They showed that the CAMs at high resolution were accurate in highlighting interpapillary capillary loop patterns in endoscopy images, which were relatively small compared to the entire image.

#### *3.1.1.3. Gradient-weighted Class Activation Mapping (Grad-CAM)*

Selvaraju et al. (2017) introduced Gradient-weighted Class Activation Mapping (Grad-CAM), which is a generalization of CAM. Grad-CAM can work with any type of CNN to produce post hoc local explanation, whereas CAM specifically needs global average pooling. The authors also introduced guided Grad-CAM, an element-wise multiplication between guided backpropagation and Grad-CAM. Grad-CAM and Guided Grad-CAM have been used in medical image analysis. For example, Ji (2019) used Grad-CAM to show on which areas of histology lymph node sections a classifier based its decision of metastatic tissue; Kowsari et al. (2020) used it to pinpoint small bowel enteropathies on histology; and Windisch et al. (2020) used Grad-Cam to show which areas of brain MRI made the classifier decide on the presence of a tumor.

#### *3.1.1.4. Layer-wise relevance propagation (LRP)*

Bach et al. (2015) introduced layer-wise relevance propagation (LRP). LRP uses the output of the neural network, e.g. a classification score between 0 and 1, and iteratively backpropagates this throughout the network. In each iteration (i.e., each layer), LRP assigns a relevance score to each of the input neurons from the previous layers. These distributed relevance scores must equal the total relevance score of its source neuron, according to the conservation law.

LRP has been used in medical image analysis. For example, Böhle et al. (2019) used LRP for identifying regions responsible for Alzheimer's disease from brain MR images. They compared the saliency maps provided by LRP with those provided by guided backpropagation, and found that LRP was more specific in identifying regions known for Alzheimer's disease.

#### *3.1.1.5. Deep SHapley Additive exPlanations (Deep SHAP)*

Lundberg and Lee (2017) proposed a unified approach for explaining model predictions by using SHapley Additive exPlanations (SHAP). This model-agnostic approach used Shapley values (Shapley (2016)), a concept from game theory. Shapley values determine the marginal contribution of every feature to the model's output individually. A downside of Shapley values is that they are resource-intensive to compute, since they require assessment of many permutations.

By combining DeepLIFT with Shapley values, Lundberg and Lee (2017) proposed a fast method to approximate Shapley values for CNNs called Deep SHAP. Deep SHAP has been used in medical image analysis. For example, van der Velden et al. (2020) used a regression CNN to estimate the volumetric breast density from breast MRI. Deep SHAP was used to explain which parts of the image had a positive contribution  and a which parts a negative contribution to the density estimation.

#### *3.1.1.6. Trainable attention*

While many of the previously mentioned techniques highlighted what regions of the image the network focuses on, i.e. to where the attention was directed, Jetley et al. (2018) proposed a trainable attention mechanism. This trainable attention method highlighted where and in what proportion the network payed attention to input images for classification, and used this attention to further amplify relevant areas and suppress irrelevant areas.

In medical imaging, Schlemper et al. (2019) used trainable attention and introduced grid attention. The rationale behind this was that most objects of interest in medical images are highly localized. By using grid attention, the trainable attention captured the anatomical information in medical images. They demonstrated high performance for both segmentation and localization, by adding the attention gates to a UNET (Ronneberger et al. (2015)) and a variant of VGG (Simonyan and Zisserman (2014)). The attention coefficients were used to explain on which areas of the image the network focused.

### 3.1.2. Perturbation-based approaches

#### *3.1.2.1. Occlusion sensitivity*

Perturbation-based techniques perturb the input image to assess the importance of certain areas of that image for the task under consideration. Zeiler and Fergus (2014) used an occlusion sensitivity analysis to visualize which parts of the image were most important for classification. For example, they showed that an image of a dog holding a tennis ball was correctly classified by the dog's breed, except if the face of the dog was occluded, which yielded the incorrect classification 'tennis ball'.

#### *3.1.2.2. Local Interpretable Model-agnostic Explanations (LIME)*

Ribeiro et al. (2016) introduced Local Interpretable Model-agnostic Explanations (LIME). LIME provides local explanation by replacing a complex model locally with simpler models, for example by approximating a CNN by a linear model. By perturbing the input data, the output of the complex model changes. LIME uses the simpler model to learn the mapping between the perturbed input data and the change in output. The similarity of the perturbed input to the original input is used as a weight, to ensure that explanations provided by the simple models with highly perturbed inputs have less effect on the final explanation. In images, Ribeiro et al. (2016) implemented the perturbations using superpixels (Achanta et al. (2012)), rather than individual pixels, to show which regions were important for explaining a classification.

LIME has been used by several researchers in medical image analysis. For example, Malhi et al. (2019) used LIME to explain which areas in gastral endoscopy images contained bloody regions.

#### *3.1.2.3. Meaningful perturbation*

Fong and Vedaldi (2017) introduced meaningful perturbation, where they perturbed the input image to detect changes in the predictions of a trained model. Rather than using perturbations such as occlusion sensitivity that block out parts of the image, they suggested simulating naturalistic or plausible effects, leading to more meaningful perturbations, and consequently to more meaningful explanations. They opted for three types of local perturbations, namely a constant value, noise, or blurring.

Uzunova et al. (2019) stated that the perturbations proposed by Fong and Vedaldi (2017) were not suited for medical images. Replacing areas of a medical image with a constant value is implausible, and medical images naturally tend to be noisy and blurry. They proposed to replace pathological regions with a healthy tissue equivalent using a variational autoencoder (VAE). They showed that the perturbations by the VAE pinpoint pathological regions in diverse imaging studies as optical coherence tomography images of the eye (pathology consisted of intraretinal fluid, subretinal fluid, and pigment epithelium detachments), and MRI of the brain (pathology consisted of stroke lesions). Furthermore, they showed that using a VAE yielded better localization of pathology compared with using simple blurring or constant-value perturbations.

Lenis et al. (2020) used similar reasoning as Uzunova et al. (2019), and used inpainting to replace pathological regions with healthy tissue equivalents. They showed that the perturbations created by inpainting outperformed backpropagation and Grad-CAM in pinpointing masses in breast mammography and tuberculosis on chest X-rays, based on the Hausdorff distance between thresholded heatmaps derived from the saliency maps and the ground truth labels at pixel level.

### *3.1.2.4. Prediction difference analysis*

Zintgraf et al. (2017) adapted prediction difference analysis (Robnik-Šikonja and Kononenko (2008)) for generating saliency maps. If each pixel in an image is considered a feature, prediction difference analysis assigns a relevance value to each pixel, by measuring how the prediction changes if the pixel is considered unknown. Zintgraf et al. (2017) expanded this by adding conditional sampling, which means that they only analyzed pixels that are hard to predict by simply investigating neighboring pixels, and by adding multivariable analysis, which means that they analyzed patches of connected pixels instead of single pixels. They included an analysis of brain MRI of patients with HIV versus healthy controls, yielding explanation of the classifier's decision.

Seo et al. (2020) used prediction difference analysis in combination with superpixels (or supervoxels for 3D) on multiple scales. These multiscale supervoxel-based saliency maps provided explanations that the authors described as visually pleasing since they follow image edges. The saliency maps explained which regions were informative for a classifier to distinguish between Alzheimer's disease patients and normal controls.

### 3.1.3. Multiple instance learning-based approaches

Multiple instance learning can be used for visualizing explanations. In multiple instance learning, training sets consist of bags of instances (Dietterich et al. (1997)). These bags are labeled, but the instances are not. In medical image analysis, multiple instance learning can for example be done using a patch-based approach: An image represents the bag, and patches from that image represent the instances (Cheplygina et al. (2019)).

Several researchers have used this approach to pinpoint which instances in the bag are responsible for the classification. For example, Schwab et al. (2020) localized critical findings in chest X-ray using such a patch-based approach. Each image patch received a prediction, and the predictions were overlaid on the image to visualize on which areas the classifier based its decision. Araújo et al. (2020) used multiple instance learning to explain which areas of a fundus photograph were important for diabetic retinopathy. They assessed the severity of the disease using an ordinal scale with grades from 0 to 5. Using a patch-based approach, they provided visual explanation maps for each diabetic retinopathy grade.

**Table 2:** Papers that used saliency maps to provide explanation. For readability, the papers are sorted on anatomical location and only the first paper dealing with that anatomical location shows the location name. The column 'Main XAI technique used/based on' describes which visual explanation technique from Section 3.1 was used, or which technique the method in the corresponding paper is based on. When multiple visual explanation techniques have been applied, the most recent technique based on Table 1 has been noted. CAM = class activation mapping, CT = computed tomography, LIME = local interpretable model-agnostic explanations, LRP = Layer-wise relevance propagation, MRI = magnetic resonance imaging, OCT = optical coherence tomography, PET = positron emission tomography, SHAP = Shapley additive explanations.

| **Anatomical location** | **Authors (year)** | **Modality** | **Main XAI technique used/based on** |
|---|---|---|---|
| Bladder | Woerl et al. (2020) | Histology | CAM |
| Brain | A. Ahmad et al. (2019) | MRI | CAM |
| | Baumgartner et al. (2018) | MRI | CAM |
| | Böhle et al. (2019) | MRI | LRP |
| | Ceschin et al. (2018) | MRI | CAM |
| | Chakraborty et al. (2020) | MRI | CAM |
| | Choi et al. (2020) | PET/CT | CAM |
| | Dang and Chaudhury (2019) | MRI | LRP |
| | Dubost et al. (2019b) | MRI | Guided backpropagation |
| | Dubost et al. (2019a) | MRI | Occlusion sensitivity |
| | Dubost et al. (2020) | MRI | Trainable attention |
| | Eitel et al. (2019) | MRI | LRP |
| | Fuchigami et al. (2020) | CT | Backpropagation |
| | Y. Gao et al. (2019) | MRI | Deconvolution |
| | K. Gao et al. (2019) | MRI | CAM |
| | Grigorescu et al. (2019) | MRI | LRP |
| | Hilbert et al. (2019) | MRI | Grad-CAM |
| | Kim and Ye (2020) | MRI | Grad-CAM |
| | Kubach et al. (2020) | Histology | Guided Grad-CAM |
| | Lee et al. (2019b) | CT | CAM |
| | Q. Li et al. (2019) | MRI | CAM |
| | Lian et al. (2019) | MRI | Trainable attention |
| | Liao et al. (2020) | MRI | Grad-CAM |
| | Lin et al. (2019) | Ultrasound | CAM |
| | Natekar et al. (2020) | MRI | Grad-CAM |
| | Ng et al. (2018) | MRI | CAM |
| | Pereira et al. (2018) | MRI | Grad-CAM |
| | Pominova et al. (2018) | MRI | Grad-CAM |
| | Rezaei et al. (2020) | MRI | Backpropagation |
| | Saab et al. (2019) | CT | Multiple instance learning |
| | Seo et al. (2020) | MRI | Prediction difference analysis |
| | Shahamat and Saniee Abadeh (2020) | MRI | Occlusion sensitivity |
| | Shinde et al. (2019a) | MRI | CAM |
| | Shinde et al. (2019b) | MRI | CAM |
| | Z. Tang et al. (2019) | Histology | Grad-CAM |
| | X Wang et al. (2020) | MRI | Guided backpropagation |
| | Wei et al. (2019) | MRI | Backpropagation |
| | Windisch et al. (2020) | MRI | Grad-CAM |
| | B. Xie et al. (2020) | Ultrasound | Grad-CAM |
| | H. Xu et al. (2019) | MRI | Trainable attention |
| | H. Xu et al. (2019) | MRI | LRP |
| | Ye et al. (2019) | CT | Grad-CAM |
| | Zintgraf et al. (2017) | MRI | Prediction difference analysis |
| Breast | Akselrod-Ballin et al. (2019) | X-ray | Meaningful perturbation |
| | El Adoui et al. (2020) | MRI | Grad-CAM |
| | Gecer et al. (2018) | Histology | Occlusion sensitivity |
| | Huang et al. (2020) | X-ray | CAM |
| | C. Kim et al. (2020) | Ultrasound | CAM |
| | Lee and Nishikawa (2019) | X-ray | CAM |
| | Luo et al. (2019) | MRI | CAM |
| | Maicas et al. (2019) | MRI | Multiple instance learning |
| | Obikane and Aoki (2020) | Histology | Grad-CAM |
| | Papanastasopoulos et al. (2020) | MRI | Integrated gradient |
| | Qi et al. (2019) | Ultrasound | CAM |
| | van der Velden et al. (2020) | MRI | SHAP |
| | H. Wang et al. (2018) | X-ray | Trainable attention |
| | Xi et al. (2019) | X-ray | CAM |
| | Yang et al. (2019) | Histology | Trainable attention |
| | Yi et al. (2019) | X-ray | CAM |
| | L.-Q. Zhou et al. (2020) | Ultrasound | CAM |
| Cardiovascular | Candemir et al. (2020) | CT | Grad-CAM |
| | Cong et al. (2019) | X-ray | Grad-CAM |
| | Gessert et al. (2019) | OCT | Guided backpropagation |
| | Huo et al. (2019) | CT | Grad-CAM |
| | Patra and Noble (2020) | Ultrasound | Grad-CAM |
| | de Vos et al. (2019) | CT | Deconvolution |
| Chest | Ausawalaithong et al. (2018) | X-ray | CAM |
| | Brunese et al. (2020) | X-ray | Grad-CAM |
| | B. Chen et al. (2019) | X-ray | Grad-CAM |
| | Dunnmon et al. (2019) | X-ray | CAM |
| | Guo et al. (2020) | CT | CAM |
| | He et al. (2017) | Histology | Grad-CAM |
| | Hosny et al. (2018) | CT | Grad-CAM |
| | Huang and Fu (2019) | X-ray | CAM |
| | Humphries et al. (2020) | CT | Grad-CAM |
| | Khakzar et al. (2019) | X-ray | CAM |
| | Ko et al. (2020) | CT | Grad-CAM |
| | Kumar et al. (2019a) | CT | CAM |
| | Lei et al. (2020) | CT | CAM |
| | Z. Li et al. (2019) | X-ray | Multiple instance learning |
| | H. Liu et al. (2019) | X-ray | CAM |
| | Mahmud et al. (2020) | X-ray | Grad-CAM |
| | R. Paul et al. (2020) | CT | Grad-CAM |
| | Pesce et al. (2019) | X-ray | Trainable attention |
| | Philbrick et al. (2018) | CT | Grad-CAM |
| | Qin et al. (2020) | PET/CT | Grad-CAM |
| | Rajaraman et al. (2019) | X-ray | LIME |
| | Rajpurkar et al. (2018) | X-ray | CAM |
| | Schwab et al. (2020) | X-ray | Multiple instance learning |
| | Sedai et al. (2018) | X-ray | CAM |
| | Singla et al. (2018) | CT | Trainable attention |
| | R. Tang et al. (2019) | CT | CAM |
| | Tang et al. (2020) | X-ray | CAM |
| | Teramoto et al. (2019) | Histology | Grad-CAM |
| | van Sloun and Demi (2019) | Ultrasound | Grad-CAM |
| | K. Wang et al. (2019) | X-ray | CAM |
| | R. Xu et al. (2019) | CT | Grad-CAM |
| | H. Y. Paul et al. (2020) | X-ray | CAM |
| | Zhu and Ogino (2019) | CT | SHAP |
| Dental | Vila-Blanco et al. (2020) | X-ray | Grad-CAM |
| Eye | M. Ahmad et al. (2019) | Fundus photography | CAM |
| | Araújo et al. (2020) | Fundus photography | Multiple instance learning |
| | Costa et al. (2019) | Fundus photography | Multiple instance learning |
| | Jang et al. (2018) | Fundus photography | Guided Grad-CAM |
| | Jiang et al. (2019) | Fundus photography | CAM |
| | M. Kim et al. (2019) | Fundus photography | Grad-CAM |
| | Kumar et al. (2019b) | Fundus photography | CAM |
| | L. Li et al. (2019) | Fundus photography | Trainable attention |
| | Liao et al. (2019) | Fundus photography | CAM |
| | C. Liu et al. (2019) | Fundus photography | CAM |
| | Martins et al. (2020) | Fundus photography | Grad-CAM |
| | Meng et al. (2020) | Fundus photography | Grad-CAM |
| | Narayanan et al. (2020) | Fundus photography | CAM |
| | Perdomo et al. (2019) | OCT | CAM |
| | Quellec et al. (2020) | Fundus photography | Backpropagation |
| | Shen et al. (2020) | Fundus photography | CAM |
| | Thakoor et al. (2019) | OCT | Grad-CAM |
| | Tu et al. (2020) | Fundus photography | CAM |
| | Wang et al. (2020) | OCT | Grad-CAM |
| | Xi Wang et al. (2020) | CT | CAM |
| | X. Wang et al. (2019) | Fundus photography | CAM |
| | Zhang et al. (2019) | Fundus photography | Grad-CAM |
| | K. Zhou et al. (2020) | OCT | CAM |
| Female reproductive system | M. Gupta et al. (2020) | Histology | Grad-CAM |
| | GV and Reddy (2019) | Histology | Grad-CAM |
| | Sun et al. (2020) | Histology | CAM |
| Gastrointestinal | X. Chen et al. (2019) | CT | Grad-CAM |
| | Everson et al. (2019) | Endoscopy | CAM |
| | García-Peraza-Herrera et al. (2020) | Endoscopy | CAM |
| | Heinemann et al. (2019) | Histology | CAM |
| | Itoh et al. (2020) | Endoscopy | Grad-CAM |
| | Kiani et al. (2020) | Histology | CAM |
| | Korbar et al. (2017) | Histology | Grad-CAM |
| | Kowsari et al. (2020) | Histology | Grad-CAM |
| | Jeong Hyun Lee et al. (2020) | Ultrasound | Backpropagation |
| | Malhi et al. (2019) | Endoscopy | LIME |
| | Rajpurkar et al. (2020) | CT | Grad-CAM |
| | Shapira et al. (2020) | CT | Multiple instance learning |
| | Wang et al. (2020) | MRI | Grad-CAM |
| | S. Wang et al. (2019) | Endoscopy | CAM |
| | Wickstrøm et al. (2020) | Endoscopy | Guided backpropagation |
| | Yan et al. (2020) | Histology | CAM |
| | Zhu et al. (2020) | Histology | Trainable attention |
| Lymph nodes | Ji (2019) | Histology | Grad-CAM |
| Musculoskeletal | Bien et al. (2018) | MRI | CAM |
| | Chang et al. (2020) | MRI | CAM |
| | Cheng et al. (2019) | X-ray | Grad-CAM |
| | V. Gupta et al. (2020) | X-ray | Grad-CAM |
| | Jamaludin et al. (2017) | MRI | Guided backpropagation |
| | Y. Kim et al. (2020) | X-ray | Backpropagation |
| | Paul et al. (2019) | X-ray | CAM |
| | Zhang et al. (2020) | X-ray | Grad-CAM |
| | Zhao et al. (2018) | X-ray | CAM |
| | von Schacky et al. (2020) | X-ray | Grad-CAM |
| Prostate | Silva-Rodríguez et al. (2020) | Histology | CAM |
| | Yang et al. (2017) | MRI | CAM |
| Skin | Barata et al. (2020) | Dermatoscopy | Trainable attention |
| | Bian et al. (2019) | Photography | Backpropagation |
| | W. Li et al. (2020) | Dermatoscopy | CAM |
| | X. Li et al. (2019) | Photography | Prediction difference analysis |
| | Y. Xie et al. (2020) | Photography | CAM |
| | Y. Yan et al. (2019) | Dermatoscopy | Trainable attention |
| | Young et al. (2019) | Dermatoscopy | SHAP |
| | Zunair and Hamza (2020) | Photography | Grad-CAM |
| Skull | Y. Kim et al. (2019) | X-ray | CAM |
| Thyroid | Jeong Hoon Lee et al. (2020) | CT | Grad-CAM |
| | J. Wang et al. (2019) | Ultrasound | Attention |
| | Wang et al. (2020) | Ultrasound | CAM |
| Multiple | Chan et al. (2019) | Histology | Grad-CAM |
| | Huang and Chung (2019) | Histology | CAM |
| | Hägele et al. (2020) | Histology | LRP |
| | Kermany et al. (2018) | Multiple | Occlusion sensitivity |
| | I. Kim et al. (2019) | Multiple | CAM |
| | Langner et al. (2019) | MRI | Grad-CAM |
| | Meng et al. (2019) | Ultrasound | Trainable attention |
| | Schlemper et al. (2019) | CT | Trainable attention |
| | Tang (2020) | Multiple | CAM |
| | Upadhyay and Banerjee (2020) | Multiple | Grad-CAM |

### 3.2. Textual explanation

Textual explanation is a form of XAI that adds textual descriptions to the model. Such descriptions include relatively simple characteristics (e.g. ‘spiculated mass’), up to entire medical reports. We will describe three types of textual explanation: image captioning, image captioning with visual explanation, and testing with concept attribution.

An overview of papers using textual explanation in medical imaging is shown in Table 3.

#### 3.2.1. Image captioning

Vinyals et al. (2015) provided textual explanation for images using an end-to-end image captioning framework. They coupled a convolutional neural network for encoding of the image, with a recurrent neural network – specifically a long-short term memory net (LSTM) (Hochreiter and Schmidhuber (1997)) – for textual encoding. They used human-generated sentences as ground truth for training, and used the bilingual evaluation understudy (BLEU) metric for evaluation. The BLEU-metric describes the precision of word N-grams, i.e. a sequence of N words, between generated and reference sentences (Papineni et al. (2002)).

Singh et al. (2019) used an image captioning framework to provide textual explanation for chest X-rays. They used word-embedding databases Global Vectors (GloVe) (Pennington et al. (2014)) and the radiology variant RadGloVe (Zhang et al. (2018)) to train the LSTM, and used the aforementioned BLEU metric as well as variants METEOR, CIDER, and ROUGE (Banerjee and Lavie (2005); Lin (2004); Vedantam et al. (2015)). As expected, higher performance was reached in the generated radiology report when both RadGloVe and GloVe were used instead of just GloVe.

### 3.2.2. Image captioning with visual explanation

Several researchers combined image captioning with visual explanation. Zhang et al. (2017a) introduced a framework that used dual attention, both for text and for imaging. They used a similar approach as with image captioning, i.e. an encoder for the image and an LSTM for the text, but added dual attention. This facilitated high-level interactions between image and text predictions, and yielded visual attention maps corresponding with textual explanation in Histology images.

X. Wang et al. (2018) used a similar approach, and showed in their chest X-ray example that different parts of the textual explanation led to different areas of saliency mapping in the image. They showed a saliency map of the chest with multiple regions corresponding to different radiological findings.

Lee et al. (2019a) showed image captioning with visual explanation for breast mammograms. They added a visual word constraint loss to the text-generating LSTM, to ensure that the provided explanations follow the correct jargon of breast mammography reports. They showed that adding this loss aids in generating better textual explanation. Furthermore, they linked the radiology reports to visual saliency maps.

### 3.2.3. Testing with Concept Activation Vectors (TCAV)

Concept attributions provide explanation corresponding to high-level concepts that humans find easy to understand (Kim et al. (2018)). Using Testing with Concept Activation Vectors (TCAV), Kim et al. (2018) presented human-friendly linear explanations of the internal state of neural networks, yielding global explanation of the networks in terms of human-understandable concepts. These concepts can be provided after training of the model as a post hoc analysis. The TCAV algorithm uses user-defined sets of examples of a concept and of random non-concept examples. Such a concept might be ‘stripes’ to assess whether an image contained a zebra, or ‘spiculated mass’ to assess whether an image contained a cancer. TCAV quantified the sensitivity of a trained model to such concepts using concept activation vectors (CAVs). The response of test cases to these CAVs was then used to measure the sensitivity to that concept. The authors showed feasibility of TCAV on a medical image processing example, by relating physician annotations such as ‘microaneurysm’ to diabetic retinopathy in fundus imaging.

Clough et al. (2019) identified cardiac disease in cine-MRI by classifying the latent space of a VAE. They used TCAV to show which clinically known biomarkers were related to cardiac disease in their model. Furthermore, they reconstructed images with low peak ejection rate – a characteristic that might be related to cardiac disease – by adding the CAV to the latent space.

Graziani et al. (2020) expanded on TCAV by introducing regression concept vectors. The main addition was that, while TCAV models are binary by indicating the presence or absence of a concept, regression concept vectors model continuous-valued measures of a concept. This can be useful when investigating a continuous concept such as tumor size. Graziani et al. (2020) showed that by using regression concept vectors, they could for example explain why the network classified one area of a breast histopathology image as cancer and another as healthy: Both areas of the image scored high on the concept ‘contrast’, but the concept ‘nuclei area’, referring to a clinically used system for evaluating cell size, was different between healthy and cancerous regions.

### 3.2.4. Other textual explanation techniques

Shen et al. (2019) used what they called a hierarchical semantic CNN to predict malignancy of lung nodules on CT. They classified five textual descriptions of image characteristics representative of lung nodule malignancy that are typically assessed by a radiologist. The task of finding textual descriptions was combined with the main task of classifying lung nodule malignancy. Although their hierarchical semantic CNN did not significantly outperform a normal CNN in predicting nodule malignancy, the method did provide human-interpretable characteristics of the nodules.

**Table 3:** Papers that provide textual explanation. For readability, the papers are sorted on anatomical location and only the first paper dealing with that anatomical location shows the location name. The column 'Main XAI technique used/based on' describes which textual explanation technique from Section 3.2 was used, or which technique the method in the corresponding paper is based on. CT = computed tomography, TCAV = testing with concept activation vectors

| **Anatomical location** | **Authors (year)** | **Modality** | **Main XAI technique used/based on** |
|---|---|---|---|
| Bladder | Zhang et al. (2017b) | Histology | Image captioning with visual explanation |
| Breast | S. T. Kim et al. (2019) | X-ray | Image captioning with visual explanation |
| | Lee et al. (2019a) | X-ray | Image captioning with visual explanation |
| | Sun et al. (2019) | X-ray | Image captioning |
| Cardiovascular | Clough et al. (2019) | MRI | TCAV |
| Chest | Gasimova (2019) | X-ray | Image captioning |
| | Kashyap et al. (2020) | X-ray | Image captioning with visual explanation |
| | C. Y. Li et al. (2019) | X-ray | Image captioning with visual explanation |
| | Nunes et al. (2019) | X-ray | Image captioning with visual explanation |
| | Rodin et al. (2019) | X-ray | Image captioning with visual explanation |
| | Shen et al. (2019) | CT | Other textual explanation |
| | Singh et al. (2019) | X-ray | Image captioning |
| | Spinks and Moens (2019) | X-ray | Image captioning |
| | Tian et al. (2019) | X-ray | Image captioning |
| | X Wang et al. (2019) | X-ray | Image captioning with visual explanation |
| | Wu et al. (2018) | CT | TCAV |
| | K. Yan et al. (2019) | CT | Other textual explanation |
| | S. Yang et al. (2020) | X-ray | Image captioning |
| | Yin et al. (2019) | X-ray | Image captioning |
| | Yuan et al. (2019) | X-ray | Image captioning with visual explanation |
| Eye | Kim et al. (2018) | Fundus photography | TCAV |
| Female reproductive system | Ma et al. (2018) | Histology | Image captioning with visual explanation |
| Gastrointestinal | Tian et al. (2018) | CT | Image captioning with visual explanation |
| Kidney | Maksoud et al. (2019) | Histology | Image captioning |
| Musculoskeletal | Koitka et al. (2020) | X-ray | Image captioning |
| Multiple | Allaouzi et al. (2018) | Multiple | Image captioning |
| | Graziani et al. (2020) | Multiple | TCAV |
| | Jing et al. (2018) | Multiple | Image captioning with visual explanation |
| | Pelka et al. (2019) | X-ray | Image captioning |
| | Zeng et al. (2020) | Multiple | Image captioning |

### 3.3. Example-based explanation

Example-based explanation is an XAI technique that provides examples relating to the data point that is currently being analyzed. This can be useful when trying to explain why a model came to a decision, and is related to how humans reason. For example, when a pathologist examines a biopsy of a patient that shows similarity with an earlier patient examined by the pathologist, the clinical decision may be enhanced by knowing the assessment of that earlier biopsy.

Example-based explanation often optimizes the hidden layers deep in the neural network (i.e., the latent space) in such a way that similar points are close to each other in this latent space, while dissimilar points are further away in the latent space.

An overview of papers using example-based explanation in medical imaging is shown in Table 4.

#### 3.3.1. Triplet network

Several papers provided example-based explanation using a triplet network (Hoffer and Ailon (2015)). A triplet network consists of three identical networks with shared parameters. By feeding these networks three input samples, the network calculates two values consisting of the $L_2$ distances between the representations in the latent space (i.e., embedded representations) of these input samples. This allows learning of useful representations by unsupervised comparison of samples. When analyzing a data point, inspection of neighbors in this embedded representation will provide examples of data points that are similar to the data point that is being analyzed, which can provide explanation why the network came to its output.

Peng et al. (2019) used example-based explanation in colorectal cancer histology. They first trained a CNN using a triplet loss, hashing, and *k* hard-negatives to learn an embedding that preserves similarity. In testing, a coarse-to-fine search yielded the 10 nearest examples from a testing database related to the input image. This provided explanation on which images similar to the image that was being analyzed the network based a decision.

Yan et al. (2018) utilized a radiological picture archiving and communication systems (PACS) to extract 32000 clinically relevant lesions from the entire body. To learn relevant lesion embeddings, they trained a triplet network with three supervision cues: lesion size, lesion anatomical location (e.g. lung, liver, or kidney), and relative coordinate of the lesion in the body. These embeddings showed good separation based on anatomical location (e.g., liver lesions were separated from lung lesions), and could accurately retrieve example-based explanation from a test set.

Codella et al. (2018) also used a triplet loss but combined it with global average pooling, the technique used in CAM. Consequently, they could not only extract example-based explanation, but they also provided query activation maps and search result activation maps. In other words, a visual explanation showed which region of the input image the network used to generate the example-based explanation. They demonstrated this technique in dermatology images of melanoma.

### 3.3.2. Influence functions

Wei Koh and Liang (2017) proposed to use influence functions to explain on which inputs from a training set the model based its decision. They did so by investigating what would happen in case an input from the training set would not be available or would be changed. Since it is expensive to assess this by perturbation, they provided an efficient approximation using influence functions (Cook and Weisberg (1980)).

C. J. Wang et al. (2019) used influence functions to explain which classifications of liver lesions on multiphase MRI were associated with which radiological characteristics. This global explanation provided insight into the neural network's behavior. For example, the class 'benign cyst' was most often associated with the radiological finding 'thin-walled mass'. Since the network did not only output the class label but also the corresponding radiological characteristics, this explanation could enhance user trust in the output of the network.

### 3.3.3. Prototypes

C. Chen et al. (2019) proposed to use typical examples as explanation (i.e., prototypes), which they described as 'this-looks-like-that'. The method reflected case-based reasoning that humans perform. For example, when a person explains why a picture contains a car, they can internally reason that this is a car because it looks like a car they have seen before. A prototype layer was added to the neural network, which grouped training inputs according to their classes in the latent space. A prototype was picked for each class, consisting of a typical example of that class. During testing, the method utilized parts of the test image that resembled these trained prototypes. The output was a weighted combination of the similarities to these prototypes. Hence, the explanation was an actual computation of the model, not a post hoc approximation.

Uehara et al. (2019) used prototypes to explain why a neural network classified patches of histology images as cancer or as not-cancer. The network was able to identify on which parts of the image it based its decision, and to what extent these parts of the image were similar to prototypical examples learned from the training set.

### 3.3.4. Examples from the latent space

Sarhan et al. (2019) proposed learning disentangled representations of the latent space using a residual adversarial VAE with a total correlation constraint. This adversarial VAE enhanced the fidelity of the reconstruction and provided more detailed descriptions of underlying generative characteristics of the data. When analyzing reconstructions by traversing through the latent space, they showed that their method yielded reconstructions that were more true to human-interpretable concepts such as lesion size, lesion eccentricity, and skin color compared with a regular VAE.

Biffi et al. (2020) provided a framework for explainable anatomical shape analysis using a ladder VAE (Sønderby et al. (2016)). They coupled this ladder VAE with a multi-layered perceptron, enabling the network to train end-to-end for classification tasks. By doing this, the highest level of the latent space was enforced to be low-dimensional (2D or 3D), which meant that these learned latent spaces could be directly visualized without the need of further dimensionality reduction after training. They provided dataset-level explanation using these low-dimensional latent spaces to visualize differences in shape for hypertrophic cardiomyopathy versus healthy controls on cardiac MRI, and for Alzheimer's disease versus healthy controls on brain MRI by visualizing the shape of the hippocampus.

Silva et al. (2018) proposed example-based explanation that showed similar and dissimilar cases foraesthetic results of breast surgery on photos, and for skin images on dermoscopy. They identified these examples using a nearest neighbor search in latent space: The nearest neighbor of the same class was considered the most similar case, and the nearest neighbor of the other class was considered the most dissimilar case. Their explanation also included rule extraction from meta-features (e.g. the color of a skin lesion or the visibility of scars). They proposed three criteria to measure the validity of the rule-extracted explanation, namely: 1) completeness, i.e. the explanation should be general enough to be applied to more than one observation; 2) correctness, i.e. if the explanation itself was considered a model, it should correctly identify which class it belongs to; and 3) compactness, i.e. the explanation should be succinct.

In later work, Silva et al. (2020) combined example-based explanation with saliency mapping. First, they trained a baseline CNN to classify chest X-rays into pleural effusion versus non-pleural effusion. After that, the CNN was fine-tuned on saliency maps. In testing, a nearest neighbor search between the latent space of the test image and a curated 'catalogue' set of images was performed. Adding the saliency map yielded more consistent examples than extracting examples without the saliency map (i.e., the baseline CNN).

Sabour et al. (2017) showed that by replacing the scalar feature maps from convolution neural networks by vectorized representations (i.e., capsules), they were able to encode high-level features of images. Capsules were basically subcollections of neurons in a layer. These were linked to subcollections of neurons in subsequent layers, forming a capsule network. This capsule network was optimized using dynamic routing. In short, higher level capsules were activated if their corresponding lower-level capsules are active. This correspondence was described by routing coefficients, which summed to one for each capsule. The coefficients were iteratively (i.e., dynamically) updated when the capsule network received new input data. For the MNIST digits dataset, Sabour et al. (2017) found that these capsules learn human-interpretable features such as scale, thickness, and skew.

LaLonde et al. (2020) used capsules for lung cancer diagnosis, while also predicting visual attributes such as sphericity, lobulation, and texture. Since these visual attributes were not necessarily mutually exclusive, as was the case in MNIST (a digit cannot be a two and a nine at the same time), they adapted the dynamic routing algorithm accordingly. Specifically, the routing coefficients did not have to sum to one in their implementation. LaLonde et al. (2020) showed that their implementation was indeed able to predict these visual attributes as well as lung nodule malignancy.

**Table 4:** Papers that provide example-based explanation. For readability, the papers are sorted on anatomical location and only the first paper dealing with that anatomical location shows the location name. The column 'Main XAI technique used/based on' describes which example-based explanation technique from Section 3.3 was used, or which technique the method in the corresponding paper is based on. CT = computed tomography, MRI = magnetic resonance imaging.

| Anatomical location | Authors (year) | Modality | XAI technique used/based upon |
|---|---|---|---|
| Brain | Y. Li et al. (2019) | MRI | Examples from the latent space |
| Breast | Uehara et al. (2019) | Histology | Prototypes |
| Chest | LaLonde et al. (2020) | CT | Examples from the latent space |
| | Silva et al. (2020) | X-ray | Examples from the latent space |
| Gastrointestinal | Peng et al. (2019) | Histology | Triplet network |
| | C. J. Wang et al. (2019) | MRI | Influence functions |
| Skin | Codella et al. (2018) | Dermatoscopy | Triplet network |
| | Sarhan et al. (2019) | Dermatoscopy | Examples from the latent space |
| Thyroid | Chen et al. (2020) | Histology | Examples from the latent space |
| | M. Li et al. (2020) | Ultrasound | Prototypes |
| Multiple | Biffi et al. (2020) | MRI | Examples from the latent space |
| | Choudhary et al. (2019) | Histology | Triplet network |
| | Silva et al. (2018) | Multiple | Examples from the latent space |
| | Yan et al. (2018) | CT | Triplet network |
| | P. Yang et al. (2020) | Histology | Examples from the latent space with visual explanation |

# 4. Discussion

## 4.1. Overview

We have discussed 223 papers on eXplainable Artificial Intelligence (XAI) for deep learning in medical image analysis. We categorized the papers based on the XAI-frameworks proposed by Adadi and Berrada (2018) and Murdoch et al. (2019). Some trends were noticeable in the surveyed papers. The majority of the papers used post hoc explanation as contrasted with model-based explanation, i.e., the explanation was provided on a model that had already been trained, instead of being incorporated in model training. Both model-specific (e.g., specifically designed for CNNs) and model-agnostic explanation methods were used. Furthermore, most of the papers investigated provided local explanation rather than global explanation, i.e., the explanation was provided per case (e.g. per patient), rather than on a dataset-level (e.g. for all patients). Since we focus on deep learning in medical image analysis, these trends were to be expected. Most readily available XAI methods suitable for CNNs are saliency mapping techniques, which often provide post hoc, model-specific, and local explanation. Furthermore, post hoc XAI methods can be used after a neural network has been trained, making them more accessible than model-based XAI.

We categorized the papers based on anatomical location and modality of medical imaging. We found that most papers focus on chest or brain and on X-ray or MRI (Figure 3). This is comparable to what Litjens et al. (2017) found for deep learning methods in medical imaging in general.

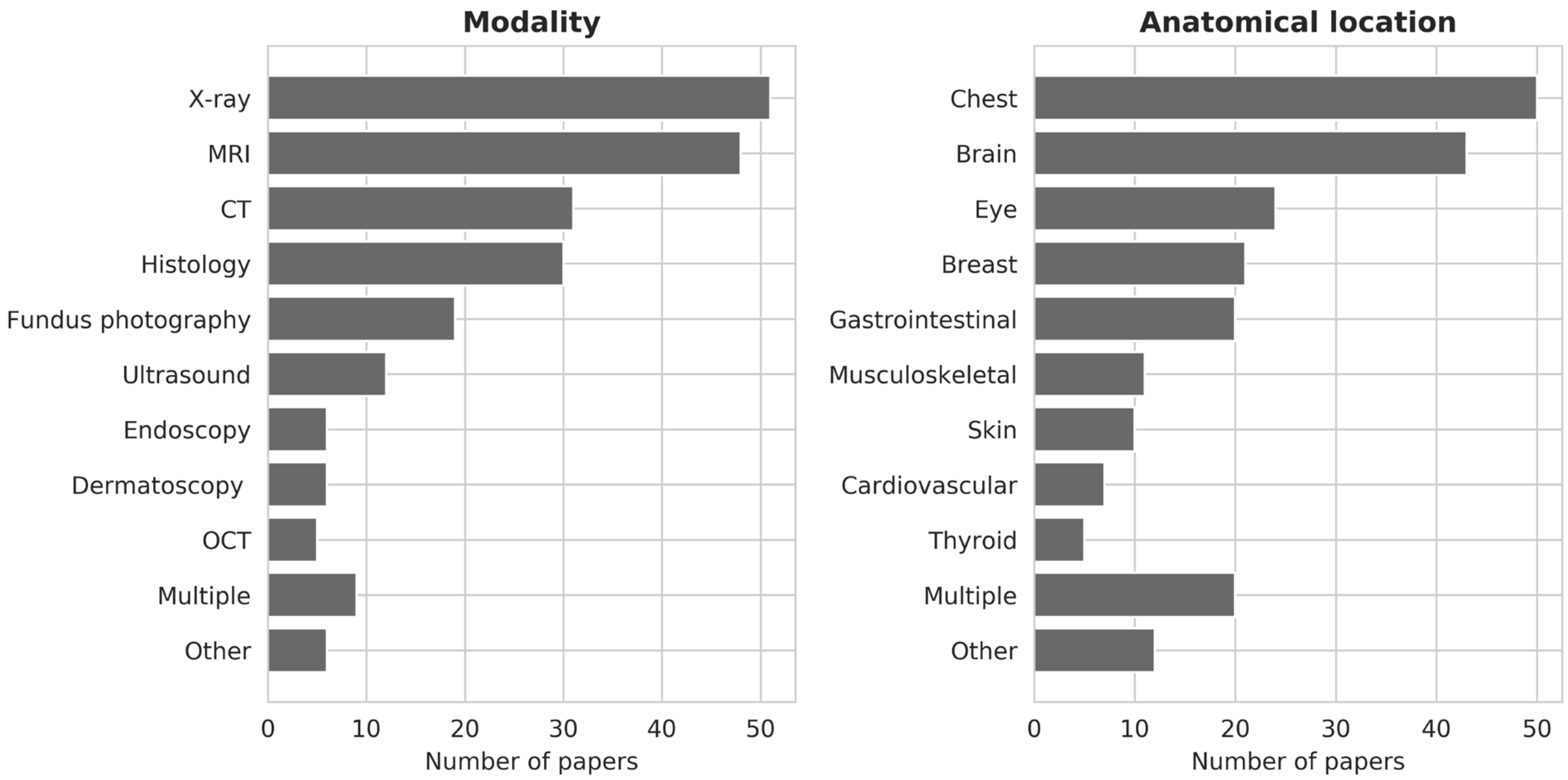

**Figure 3:** Papers included in this survey, categorized by modality (left) and anatomical location (right). Papers discussing multiple modalities or anatomical locations were grouped as 'multiple'. Modalities or anatomical locations that were used in fewer than five papers were grouped as 'other'.

## 4.2. Evaluation of XAI

We have described several XAI techniques and their applications in medical image analysis, but how does one evaluate whether an XAI technique provides good explanation? Unlike measures of performance commonly used in medical image analysis, such as accuracy, Dice coefficient, or an ROC analysis; success criteria of explanation are more difficult to define. Doshi-Velez and Kim (2017) proposed a framework for the evaluation of explainability, consisting of three evaluation methods: application-grounded evaluation, human-grounded evaluation, and functionally-grounded evaluation.

#### 4.2.1. Application-grounded evaluation

Application-grounded evaluation uses human experiments within a real application. In other words, let domain experts test the explanation. In medical image analysis this might involve a radiologist inspecting whether example-based explanations are actually good examples based on the many images the radiologist has seen in their many years of experience. The advantage of application-grounded evaluation is that it directly tests the objective that the system was built for. The disadvantage is that it is a costly evaluation.

#### 4.2.2. Human-grounded evaluation

Human-grounded evaluation uses simpler human experiments that maintain the essence of the target application. In other words, let laypersons test the explanation or a proxy of the explanation. For example, when explaining the location and size of a cancer, this might involve a crowdsourcing project where laypersons judge the quality of saliency maps. Since it uses laypersons instead of highly trained domain experts, the advantage of human-grounded evaluation is that it is less costly, while still receiving general notions of the quality of an explanation. The disadvantage is that the assessment of the quality of an explanation is a proxy of the actual quality.

#### 4.2.3. Functionally-grounded evaluation

Functionally-grounded evaluation does not use human experiments, but uses other proxies to assess the quality of the explanation. These proxies may include measurements that have already been validated using human users. In our example of explaining the location and size of a cancer, this might involve comparing the explanation with manually drawn tumor delineations of a radiologist. The advantages of functionally-grounded evaluation stated by Doshi-Velez and Kim (2017) include that they are relatively cheap to acquire. This is, however, not necessarily the case in medical image analysis, since acquiring for example manual annotations is a very resource intensive process. When these manual annotations do already exist, e.g. when using curated data from a challenge, evaluation of explanations are easily extracted, and can be automatically extracted multiple times. This can be useful, for example in the development phase of explanation methods.

### 4.3. Critique on XAI

Rudin (2019) advised caution when using black box models with explanation for high-stakes decision making. Rudin raised several issues with explaining black box models. For example, XAI may provide an explanation that is not completely faithful to what the original model computes: If the explanation explains 90% true to the model, that means that 10% is untrue (Rudin (2019)). Furthermore, an explanation may not make sense or provide enough detail to understand what the black box is doing. For example, a saliency map of the class with the highest probability may look similar to a saliency map of a class with a lower probability. Rudin therefore advices to use interpretable model-based XAI instead, such as the prototype network discussed in section 3.3.3.

Adebayo et al. (2018) investigated the robustness of several saliency mapping techniques using two tests: parameter randomization and data randomization.

The parameter randomization test compared saliency maps from a trained CNN with saliency maps from a randomly initialized untrained CNN of the same architecture. If the saliency map depended on the learned parameters of the CNN (the desired situation), the two saliency maps should have differed substantially. If the two saliency maps were similar, the saliency mapping technique was insensitive to the properties of the CNN.

The data randomization test compared saliency maps from a trained CNN with saliency maps from a CNN trained on the same dataset but with randomly imputed labels. If the saliency map depended on the data labels (the desired situation), the two saliency maps should have differed substantially. If the two saliency maps were similar, the saliency mapping technique did not depend on the relationship between images and labels.

Adebayo et al. (2018) performed these two tests for many visual explanation methods including backpropagation, guided backpropagation, and guided Grad-CAM. They showed that guided backpropagation and guided Grad-CAM showed similar saliency maps in both tests, and might be emphasizing edges. Hence, caution is advised when using such methods for visualization.

Eitel and Ritter (2019) evaluated the robustness of saliency mapping techniques in medical images over multiple training runs, specifically for the classification of Alzheimer's disease on brain MRI. They found that layer-wise relevance propagation and guided backpropagation produced the most coherent attribution maps. This was not fully in line with the results of Adebayo et al. (2018). Hence, more research on this topic in medical image analysis is desired.

### 4.4. Outlook

Since high stakes decision-making is intertwined with medicine, we are convinced that XAI will be increasingly important. We have investigated the trends, and noticed that an increasing amount of papers contain a holistic approach, combining multiple forms of explanation. Examples of such more holistic approaches include combinations of textual explanation and visual explanation (e.g. Graziani et al. (2020)), or combinations of example based explanation and visual explanation (e.g. C. J. Wang et al. (2019)).

Future directions of XAI in medical image analysis may include biological explanation. Several researchers have predicted biological processes from imaging features using deep learning. For example, Matsui et al. (2020) predicted the molecular subtype of lower-grade gliomas on multimodal brain imaging, and Zhu et al. (2019) predicted the molecular subtype luminal A of breast cancer on MRI. These analyses used a biological target to train the neural network. However, performing such analysis the other way around, for example by performing a pathway analysis on imaging phenotypes (e.g. Bismeijer et al. (2020), not deep learning), could provide interesting biological explanation.

Other directions of XAI in medical image analysis may include the link between causality and XAI. Typical medical image analysis consists of correlation rather than causation. Causality describes the relation between cause and effect, and can be mathematically described (Pearl (2009)). Current XAI techniques that aim to be free of bias such as prototypes are potentially still sensitive to differences in training population, which might hamper generalizability. Castro et al. (2020) describe how causal reasoning may be useful to assess such as biases in the data. van Amsterdam et al. (2019) show an example of eliminating bias using causality, yielding unbiased prediction of prognosis for patients with lung cancer. It would be of interest to incorporate such analyses in explanation of medical images, as Chattopadhyay et al. (2019) have done for visual explanation of MNIST data.

There is no consensus on a priori estimations for required sample size for XAI and deep learning in medical imaging in general (Balki et al. (2019)). Given the costly nature of acquiring medical imaging datasets in terms of money, time, and patient burden, it is desired to have guidelines describing what minimum sample sizes would be required for which XAI techniques.

### 4.5. Limitations

We derived our XAI framework from the frameworks of Adadi and Berrada (2018) and Murdoch et al. (2019). Other frameworks also exist, such as the framework by Kim et al. that divides XAI in pre-, during-, and post-model explanation. During- and post-model explanation are captured by our XAI framework with model-based and post hoc explanation. Pre-model explanation mainly focuses on the structure of a dataset, such as inspecting outliers. One could state that an example-based explanation that utilizes the latent distributions of a dataset could be perceived as a pre-model explanation. We have, however, not made this distinction, since in deep learning, these latent distributions are discovered by training a neural network.

We tried to be as comprehensive as possible with the inclusion of papers in our survey. However, XAI often is a technique used to support methods, and keywords are often not mentioned in the title or body of papers (Rudin (2019)). Therefore, we cannot guarantee that we covered all the work in the field. Nevertheless, we provided the search strategy in the appendix to be as transparent as possible about the selection of papers.

## 5. Conclusion

This paper surveyed 223 papers using explainable artificial intelligence (XAI) in deep-learning based medical image analysis, classified according to an XAI framework, and categorized according to anatomical location and imaging technique. The paper discussed how to evaluate XAI, current critiques on XAI, and future perspectives for XAI in medical image analysis.

## 6. Additional information

This work was partially funded by the Dutch Cancer Society (KWF) grant number: 10755. We have no conflicts of interest.

## 7. Appendix

We used the search query “(explainable deep learning OR interpretable deep learning OR XAI OR interpretable machine learning OR explainable machine learning) AND (medical imaging OR medical image analysis)” in SCOPUS. We analyzed the query results using the Active learning for Systematic Reviews toolbox. This toolbox uses active learning to sort papers from most relevant to least relevant, while being updated by user input. Furthermore, we had discussions with colleagues, and used a snowballing approach – investigating papers referenced by the included papers and papers that refer to the included papers. We read the title and the abstract of each of these papers, and browsed paper content if we were not sure whether to include the paper. In case of multiple publications by the same authors on the same subject, we chose the journal publication or the most recent publication in case of multiple conference publications. We included peer reviewed journal papers and conference proceedings. Papers up to October 2020 are included in the survey.